# USING ARTIFICIAL INTELLIGENCE TO DETECT CHEST X-RAYS WITH NO SIGNIFICANT FINDINGS IN A PRIMARY HEALTH CARE SETTING IN OULU, FINLAND

Keski-Filppula Tommi [1], Nikki Marko [2], Haapea Marianne [2], Ramanauskas Naglis [3], Tervonen Osmo [1,2]. Corresponding author: Keski-Filppula Tommi

1 - Research Unit of Medical Imaging, Physics and Technology, Faculty of Medicine, University of Oulu, Oulu, Finland.

2 - Department of Diagnostic Radiology, Oulu University Hospital, Oulu, Finland., P.O. Box 10, 90029 OYS

3 – Oxipit UAB, Vilnius, Lithuania

# ABSTRACT

*Objectives*

To assess the use of artificial intelligence-based software in ruling out chest X-ray cases, with no significant findings in a primary health care setting.

*Methods*

In this retrospective study, a commercially available artificial intelligence (AI) software was used to analyse 10 000 chest X-rays of Finnish primary health care patients. In studies with a mismatch between an AI normal report and the original radiologist report, a consensus read by two board-certified radiologists was conducted to make the final diagnosis.

*Results*

After the exclusion of cases not meeting the study criteria, 9579 cases were analysed by AI. Of these cases, 4451 were considered normal in the original radiologist report and 4644 after the consensus reading. The number of cases correctly found nonsignificant by AI was 1692 (17.7% of all studies and 36.4 of studies with no significant findings).

After the consensus read, there were nine confirmed false-negative studies. These studies included four cases of slightly enlarged heart size, four cases of slightly increased pulmonary opacification and one case with a small unilateral pleural effusion. This gives the AI a sensitivity of 99.8% (95% CI= 99.65-99.92) and specificity of 36.4 % (95% CI= 35.05-37.84) for recognising significant pathology on a chest X-ray.

*Conclusions*

AI was able to correctly rule out 36.4% of chest X-rays with no significant findings of primary health care patients, with a minimal number of false negatives that would lead to effectively no compromise on patient safety. No critical findings were missed by the software.

*Key point*

- Artificial intelligence can reliably and safely rule out significant pathology in a chest X-ray.

**INTRODUCTION**

The use of artificial intelligence (AI) and machine learning in radiology has been under investigation in recent years. In Finland, approximately 700 000 chest X-rays were performed in 2018, making it the most frequently utilised radiologic study after the dental X-ray [1]. Prior research indicates that AI can perform as well as, or in some cases, even better than, a radiologist, at recognising certain findings in a chest X-ray [2–5]. Currently, the most promising results against a radiologist are on detecting isolated findings, such as lung nodules, or pneumothoraces [2, 6]. AI can indeed perform well, compared against a radiologist, in recognising isolated findings, but few studies thus far have compared AI to a radiologist in a setting where the radiologist has access to the patient's clinical information, as would be the case in real life [7]. One practical angle of approach is to ask how AI can help a radiologist. Two recent studies demonstrated that a comprehensive, deep-learning model significantly improved the interpretation of a chest X-ray by a radiologist [8, 9] and was well received [8]. Another promising real-life use case of AI in the context of chest X-rays is triaging to reduce report turnaround times for critical findings [10, 11].

In this study, we explore one possible pathway for AI utilisation: ruling out normal chest X-ray studies in a primary health care setting, where a significant number of studies contain no significant findings to begin with. Our hypothesis is that AI can reliably recognise a significant percentage of normal chest X-rays with a very small number of false negatives. Furthermore, we hypothesise that, in a clinical setting, if the chest X-ray is interpreted as normal by AI, no radiologist report is needed unless the clinician requests it. This could help optimize healthcare resources by reducing the workload of chest X-ray reporting.

To our knowledge, one previous study with a setting like ours has been conducted. In a recent study, Dyer et al. demonstrated that a deep learning algorithm was able to classify 15% of unselected chest X-rays as normal with a precision of 97.7% and an error rate of 0.33%, removing 24.9% of radiologist-confirmed normal studies from the workflow [12].

# MATERIALS & METHODS

## *The source of data*

Oulu is a city of 209 000 inhabitants, located in the North Ostrobothnia region of Finland. There are several primary health care centres and one hospital – the Oulu University Hospital. Primary health care centres mostly take care of patients with stable chronic illnesses and acutely ill patients, who do not require hospital-level care. More severely ill patients are referred to the hospital.

Our data consist of all chest X-rays taken of primary health care patients residing in the city of Oulu, Finland, from the years 2019-2020, up to 10 000 images. The following are the exclusion criteria for chest X-rays in our study: children (under 18 years of age), pregnant women, images with technical issues (poor image quality, rotation, incompletely recorded lung fields) and other than a posterior-anterior (PA)-position of the image.

## *AI software*

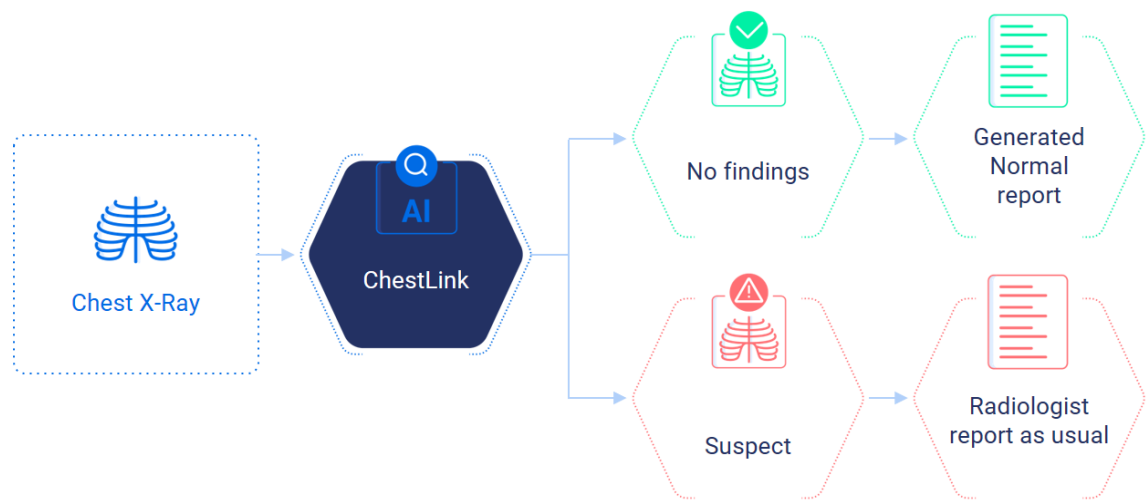

Figure 1. An illustration of the AI software used in the study (ChestLink®).

AI-based commercial software (ChestLink®, Oxipit, Vilnius, Lithuania) was used to analyse the X-rays. At the time of the data analysis, only frontal X-ray images were used, as lateral view analysis was not yet enabled in the software.

The AI software has been developed and optimised heavily towards maximising sensitivity in order to identify the studies without any abnormalities with high sensitivity. The AI has been tested with approximately one million chest X-rays, and the fraction of normal chest X-ray studies, for which a normal report is generated depending on the dataset, has ranged from 10 to 55 %.

The software analyses both frontal and lateral chest X-ray images. A normal report by the software can only be generated for the studies which are as follows: a) Erect PA (posterior-anterior radiographs) b) performed for adult patients (>= 18 years).

*Data classification and processing*

The original radiologist report, and consensus read as necessary, was considered a ground truth in our study. The content of the original radiologist report was analysed using a natural language processing (NLP) software provided by Oxipit. The software can recognise 75 radiological findings in radiologist reports and for each finding, a special rule-based system is used to capture whether it was mentioned in the radiologist report or not. The software can handle multiple ways of mentioning the given findings in a report as well as handle the negations to minimise any false positives. For non-English reports, the software utilises a translation service as a part of the text mining pipeline.

If AI assessed the study as normal, and the NLP system recognised the radiologist report as normal, then the study was considered a software-true negative / normal. If AI assessed the study as normal, but the NLP system recognised the radiologist report as abnormal, the study was considered an initial software false negative. We then excluded purely NLP-related errors, which left us with 156 cases, where there was a discrepancy between the AI and radiologist reports. Two board-certified radiologists (MN and OT) then evaluated the images, and the decision was made by consensus. We classified the study as normal if the finding originally commented in the radiologist report was considered insignificant, over-diagnosed or miscellaneous in the consensus read. There were 147 of such cases, including healed rib fractures, slight pleural thickening or pleural adhesions and small linear pulmonary atelectases.

### *Statistical analysis*

We used sensitivity, specificity, negative predictive value, positive predictive value and accuracy with their 95% confidence intervals (CI) to evaluate and represent the performance of AI. All parameters were calculated using an online MedCalc Diagnostic test evaluation calculator: https://www.medcalc.org/calc/diagnostic_test.php.

## RESULTS

### *True negatives*

Of the 9579 images, 4451 (46.4%) were reported as normal by a radiologist, as recognised by the NLP system (Figure 2). Using the original radiologist report as ground truth, AI correctly assessed 1499 images (33.6% of original radiologist negatives) as normal. When added in studies assessed normal by AI, where the NLP software incorrectly recognised the radiologist report as abnormal (N=46), and where the originally reported abnormal findings were considered insignificant in the consensus reading (N=147), there was a total number of 1692 AI true normal studies (17.7% of all studies and 36.4% of studies that a radiologist originally classified as normal).

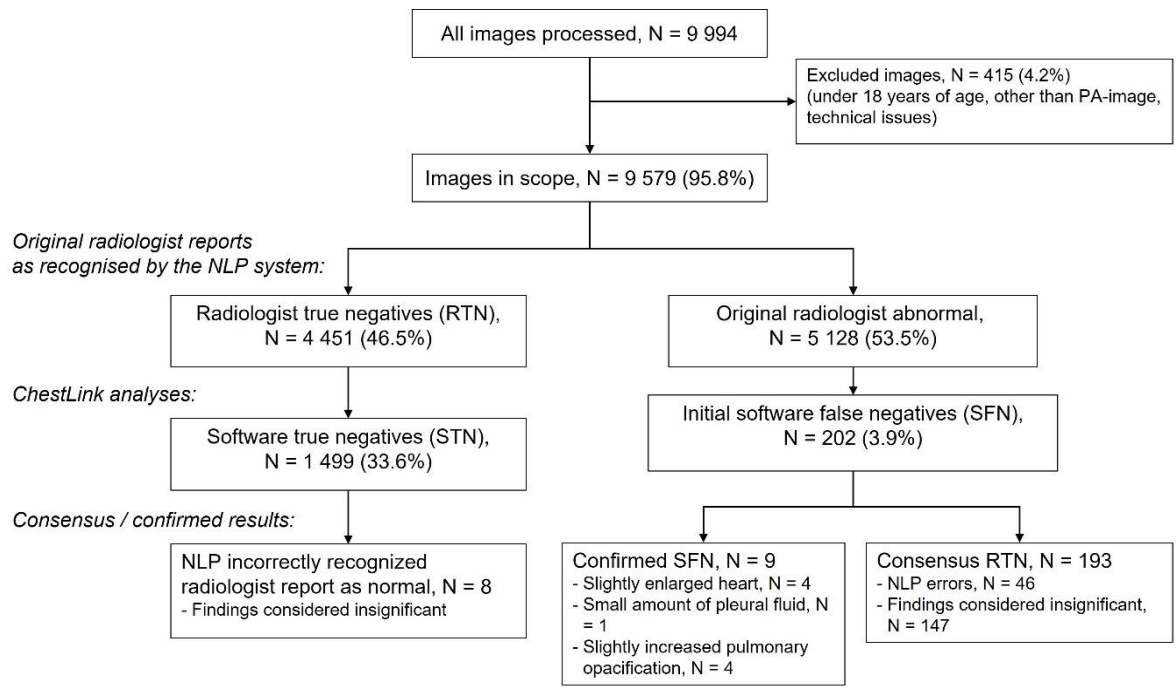

Figure 2. Flowchart representing data processing and relevant results. NLP = Natural language processing.

## *False negatives*

There were 202 cases where AI classified the case as normal, but the original radiologist report described abnormality, as recognised by the NLP system. After excluding NLP-related errors (N=46), the initial comparison between the AI and radiologist report showed 156 false negative studies. 147 of these studies were classified as containing nonsignificant, over-diagnosed or otherwise unremarkable findings in the consensus reading. These findings are presented in Figure 3.

Some of the findings, such as vertebral changes or posterior costophrenic angle blunting, could only be appreciated on lateral images. As lateral image analysis was not enabled at the time of analysis, lateral-view-only findings were classified as insignificant.

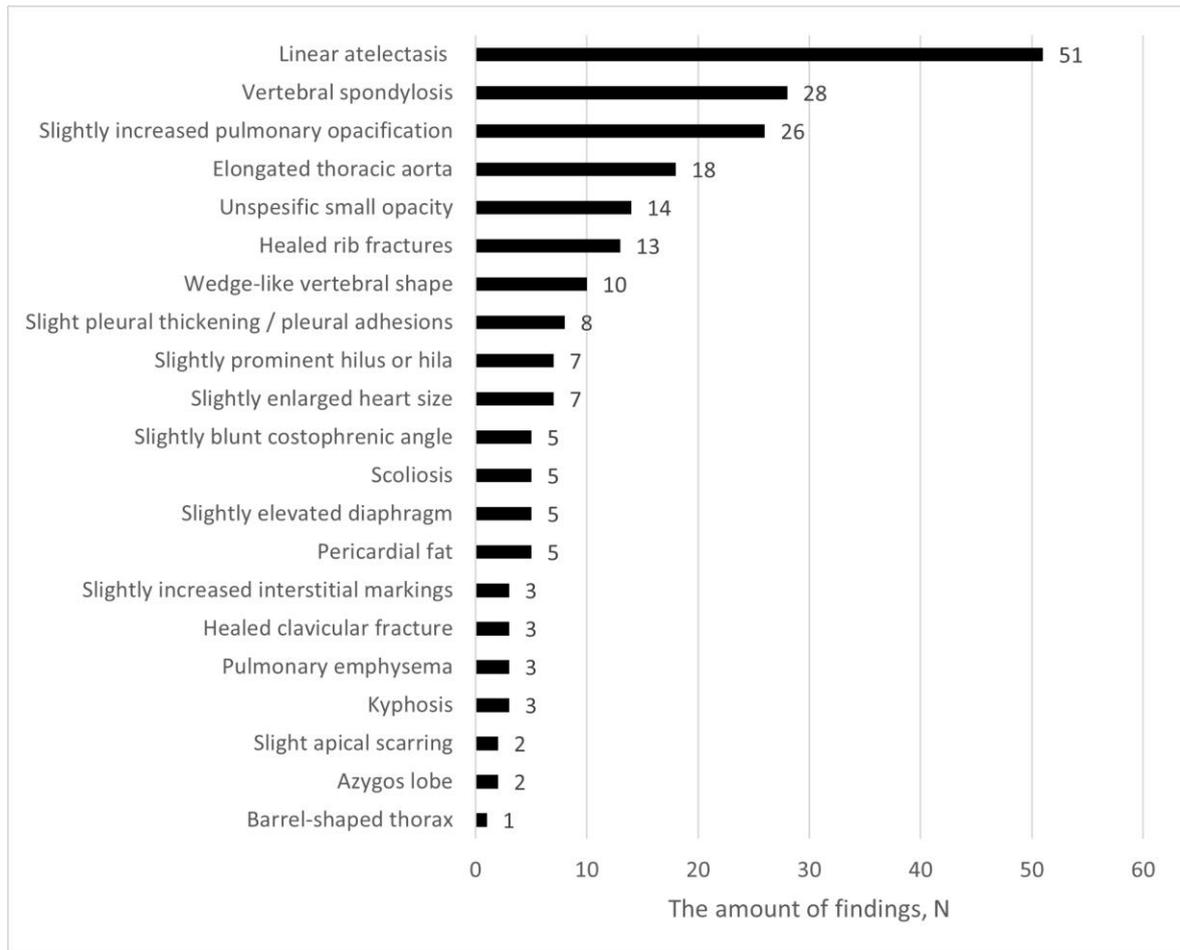

Figure 3. The findings in the chest X-ray image not depicted by AI but depicted by a radiologist and further considered as nonsignificant, over-diagnosed or otherwise unremarkable in the consensus read. The x-axis represents the absolute number of findings. It is notable that there is overlap: multiple findings can be commented in the original radiologist report of a single study. The unspecific small opacities included for example pleural plaques and mamillary shadows.

To rule out the possibility of false negative studies in the true normal group, we manually examined the radiologist reports on all AI true negative studies (N = 1499). There were 8 cases where the NLP system had incorrectly recognised the radiologist report as normal. These radiologist reports included one case with a slightly prominent right hilum, three cases of suspected peribronchovascular consolidation, one case of suspected rib fracture and three cases of suspected unspecific opacities. All these cases were classified as normal or insignificant in the consensus read. These cases are also included in Figure 3.

Of the 202 initial false negatives, nine were considered potentially significant, "confirmed false negatives". Four of these studies had a slightly increased heart size, four had slightly increased pulmonary opacification and one had a small unilateral pleural effusion. One study of each category is presented in Figure 4. Full-sized images are available in the electronic supplementary material. The other three cases of both slightly enlarged heart size and increased pulmonary opacification were comparably similar.

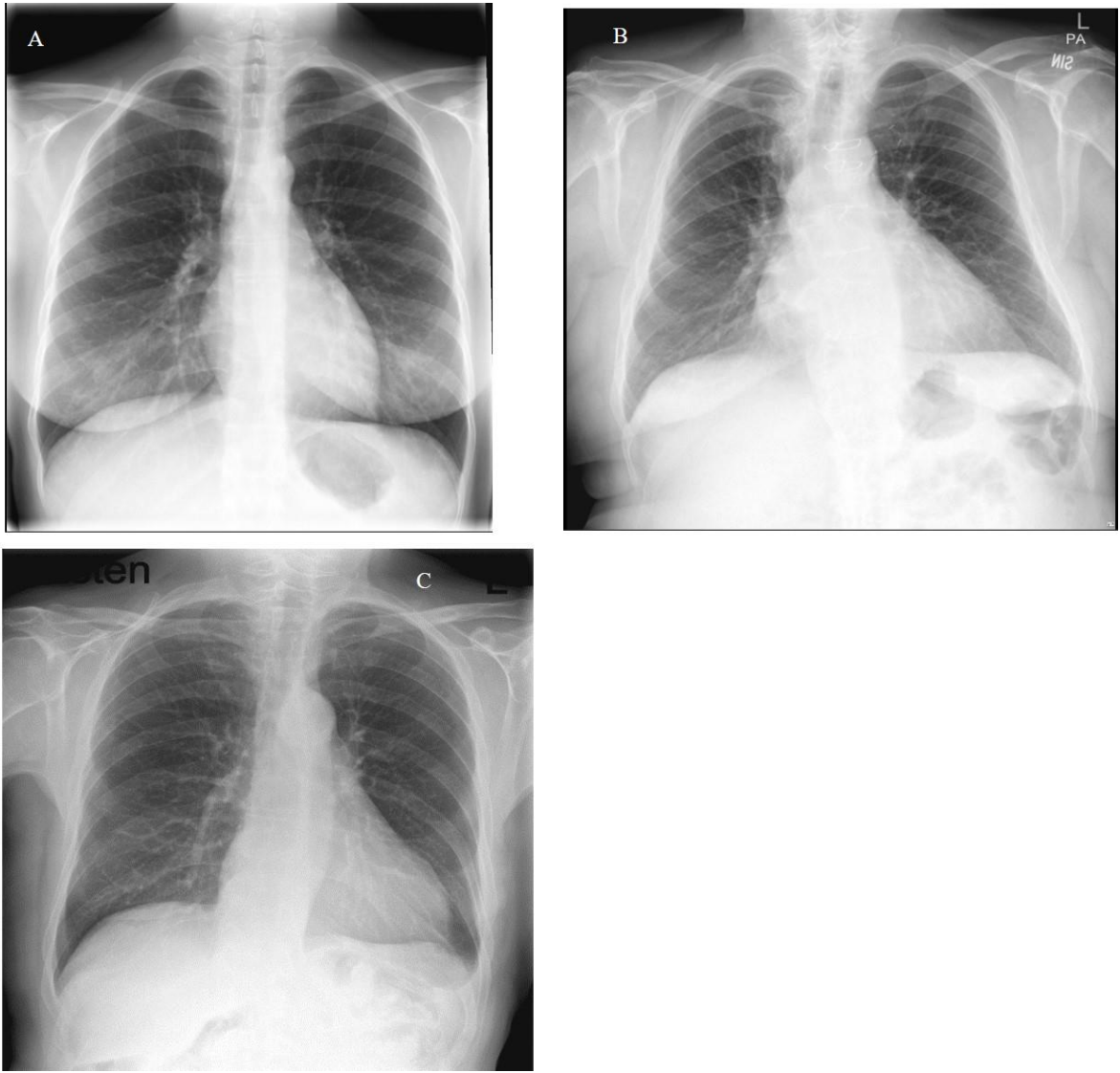

Figure 4. A representation of the confirmed false negative AI studies.

A) Bilateral basal peribronchovascular opacification, potentially indicating bronchopneumonia. B) A slightly enlarged heart size. C) A small left-sided pleural effusion.

After the consensus read, there were nine confirmed false-negative studies. This gives an AI specificity of 36.4 % (95% CI = 35.05-37.84) and a sensitivity of 99.8 % (95% CI = 99.65-99.92) for recognising significant pathology on a chest X-ray. (Table 1.)

Table 1. Performance of AI in recognising significant pathology in a chest X-ray.

| Statistic | Value (%) | 95 % CI (%) |
|---|---|---|
| Sensitivity | 99.82 | 99.65–99.92 |
| Specificity | 36.43 | 35.05–37.84 |
| Positive predictive value | 62.53 | 62.02–63.04 |
| Negative predictive value | 99.47 | 98.99–99.72 |
| Accuracy | 69.09 | 68.15–70.01 |

Percentages were rounded to two-decimal accuracy. CI = confidence interval.

## DISCUSSION

In this study, we demonstrated that AI can provide a reliable way to rule out significant pathology in chest X-rays. The number of confirmed false negatives was small. There were four cases with a slightly enlarged heart size, one case with a small amount of pleural fluid and four cases with slightly increased pulmonary opacification. The potential effect of these misdiagnoses on patient safety is negligible and none of them would lead to immediate threat to the patient. A slightly enlarged heart size is usually not an acute finding, except in the rare case of pericardial fluid, and does not require immediate treatment. A small amount of pleural fluid has little clinical relevance. The four cases with a slightly increased pulmonary opacification might indicate mild bronchopneumonia, but the finding is controversial, and interobserver variability is high among radiologist in reporting this finding. While the X-ray finding is not unequivocal, the clinical diagnosis and treatment is mostly based on the clinical condition of the patient. No lobar consolidates indicating bacterial pneumonia were missed by the software.

The number of nonsignificant findings in a chest X-ray is generally high, and the main task in the interpretation is to depict the clinically relevant ones. An AI software should be able to reliably depict the clinically relevant findings. In this study, there were many chest X-ray findings which were not depicted by AI and depicted by a radiologist, and they were further evaluated as nonsignificant by the consensus reading (Figure 3). Our approach to classifying these as nonsignificant was clear and consensus-based: in the context of primary health care, the purpose of a chest X-ray is to rule out treatable disease. The prevalence of spinal degeneration in asymptomatic people is high especially in the elderly population and does not represent disease [13]. Pulmonary linear atelectases are common, usually self-resolving, most commonly caused by hypoventilation [14]. Some of the findings, such as suspected pulmonary opacification or slightly increased interstitial markings, represent the subjectivity of chest X-ray reporting: inter-reader agreement varies considerably in a clinical setting [15–17]. In our study, the final diagnosis was made by consensus by two experienced, board-certified radiologists.

Comparing the results of this study to the study by Dyer et al. in 2021, our results are on a similar trajectory. However, in their study, 4 out of 584 high-confidence normal studies showed critical findings missed by the software: one lung nodule, two lung masses and one pulmonary consolidation. In our study, no critical findings were missed.

Using AI to rule out normal studies could provide one way to optimise the use of resources in healthcare. With the increasing volumes of CT and MRI, the radiologist workforce could thus be directed to more complicated studies. Further research in a prospective clinical setting is needed to gather real-life clinical data and use the experience of autonomous AI-based reporting of chest X-rays.

*Limitations*

The main limitation of this study is the ground truth against which the AI analysis is compared. The ground truth in our study was the original radiologist report and further consensus-read when necessary. This does represent a real-life scenario, where radiologist reports are used to guide patient care, but does potentially create uncertainty regarding borderline findings, as a definitive diagnosis based on a chest X-ray is often not possible. Also, we did not have computed tomography or follow-up to confirm the chest X-ray findings, but the evaluation was based on the consensus reading.

NLP software was used to analyse the initial radiologist reports, which may cause one source of error. To minimise the effect of NLP on the results, we manually consensus-read all the software false negative studies and checked all the original radiologist reports in the true normal group. This eliminates NLP-related errors regarding the false negative studies.

Another limitation of this study is the study population. While the study population should represent the general patient base of our primary health care centres, an accurate description of the study population is not possible, as we did not have access to patient demographic information.

This study was based on frontal chest X-ray images only, as the functionality to analyse lateral images was not available at the time of the study. This may lead to underdiagnoses of findings best appreciated in the lateral view, such as vertebral body changes or blunting of the posterior costophrenic angles.

*Conclusions*

Based on the results of this study, AI can reliably remove 36.4% of normal chest X-rays from a primary health care population data set with a minimal number of false negatives, leading to effectively no compromise on patient safety. No critical findings, such as lung masses, pneumothoraces or lobar consolidations were missed by the software.

**ACKNOWLEDGEMENTS**

The authors state that this work has not received any funding. This study was conducted in collaboration with Oxipit.

**SUPPLEMENTARY MATERIAL – AI FALSE NEGATIVE CASES**

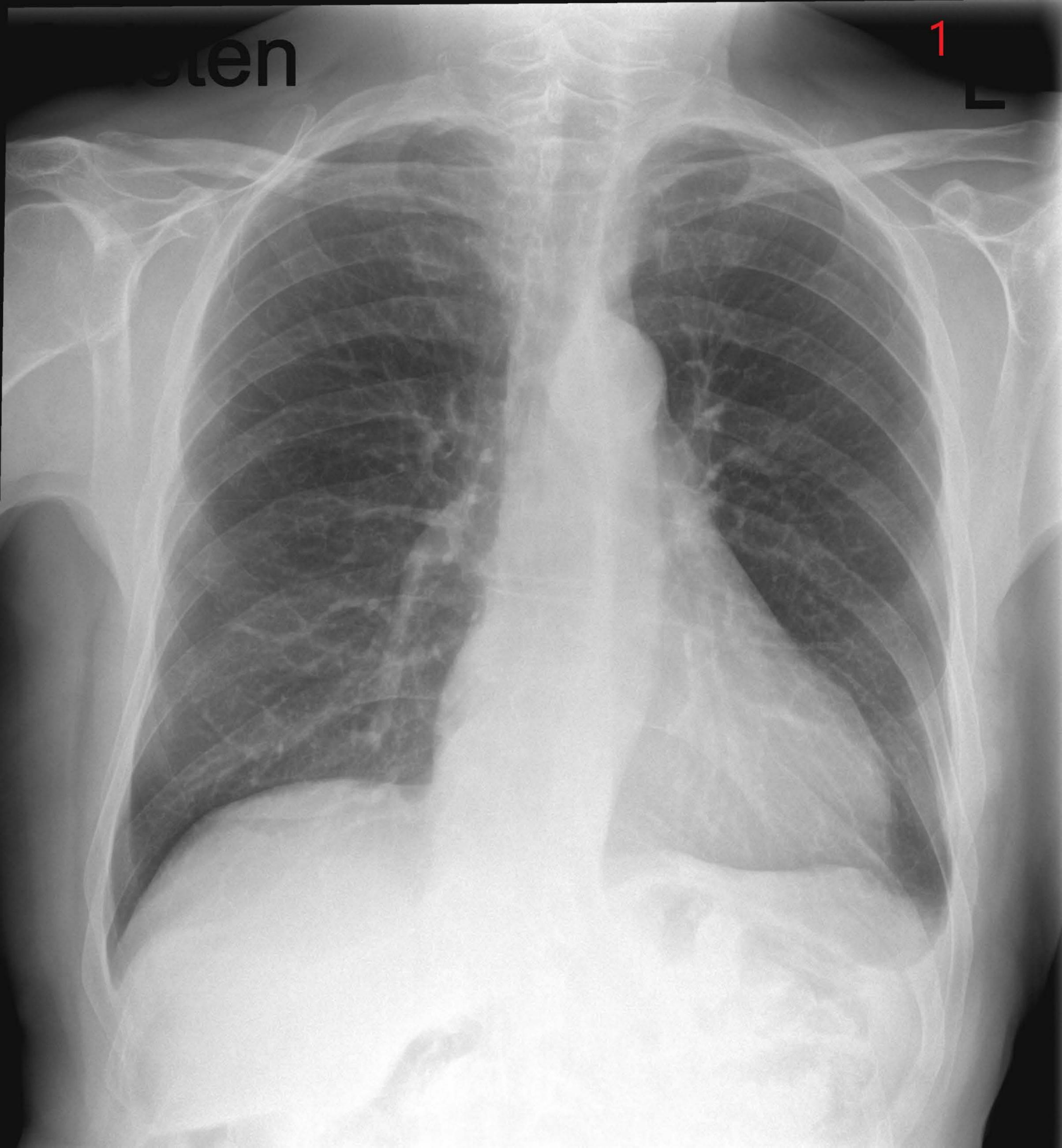

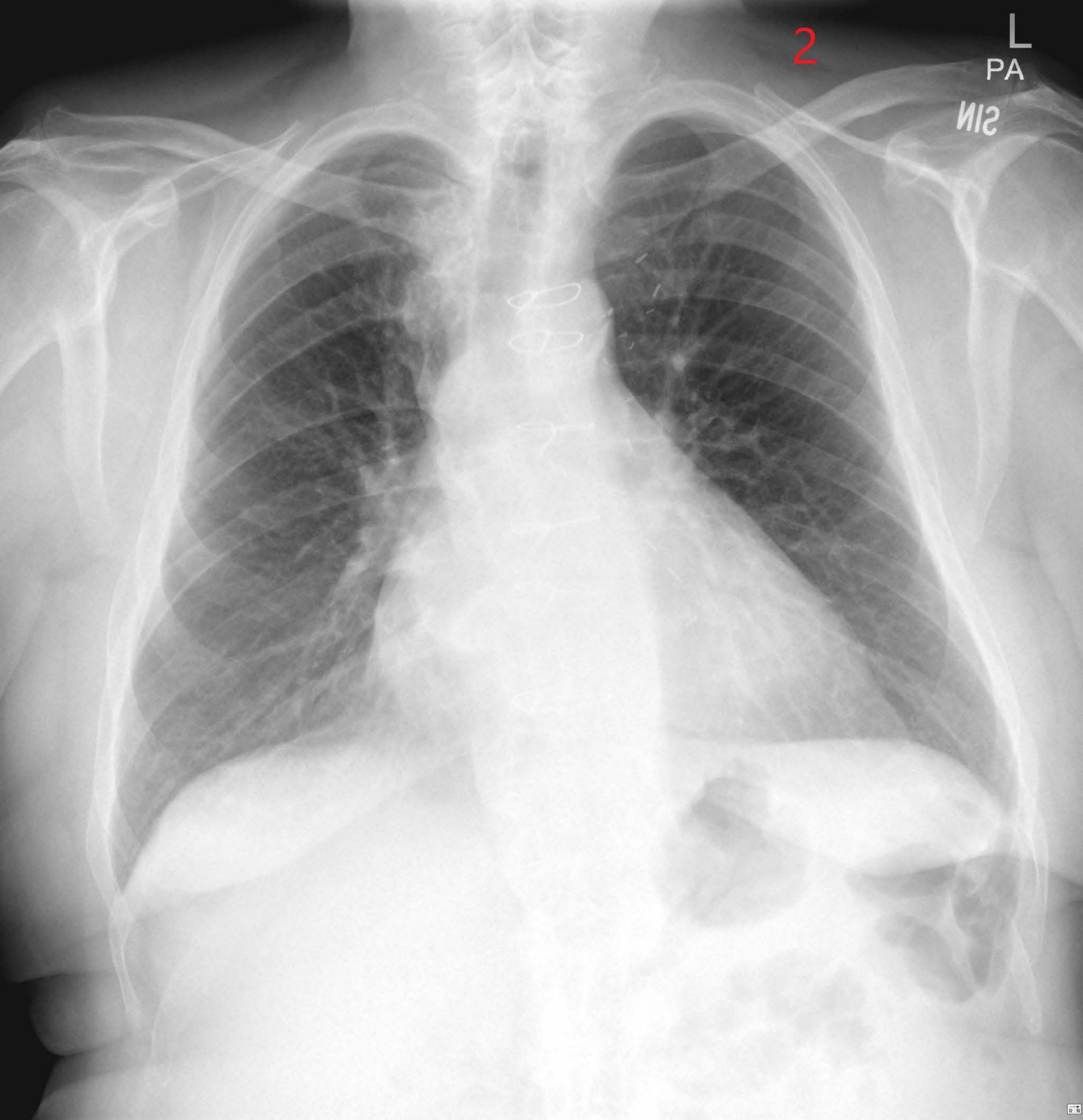

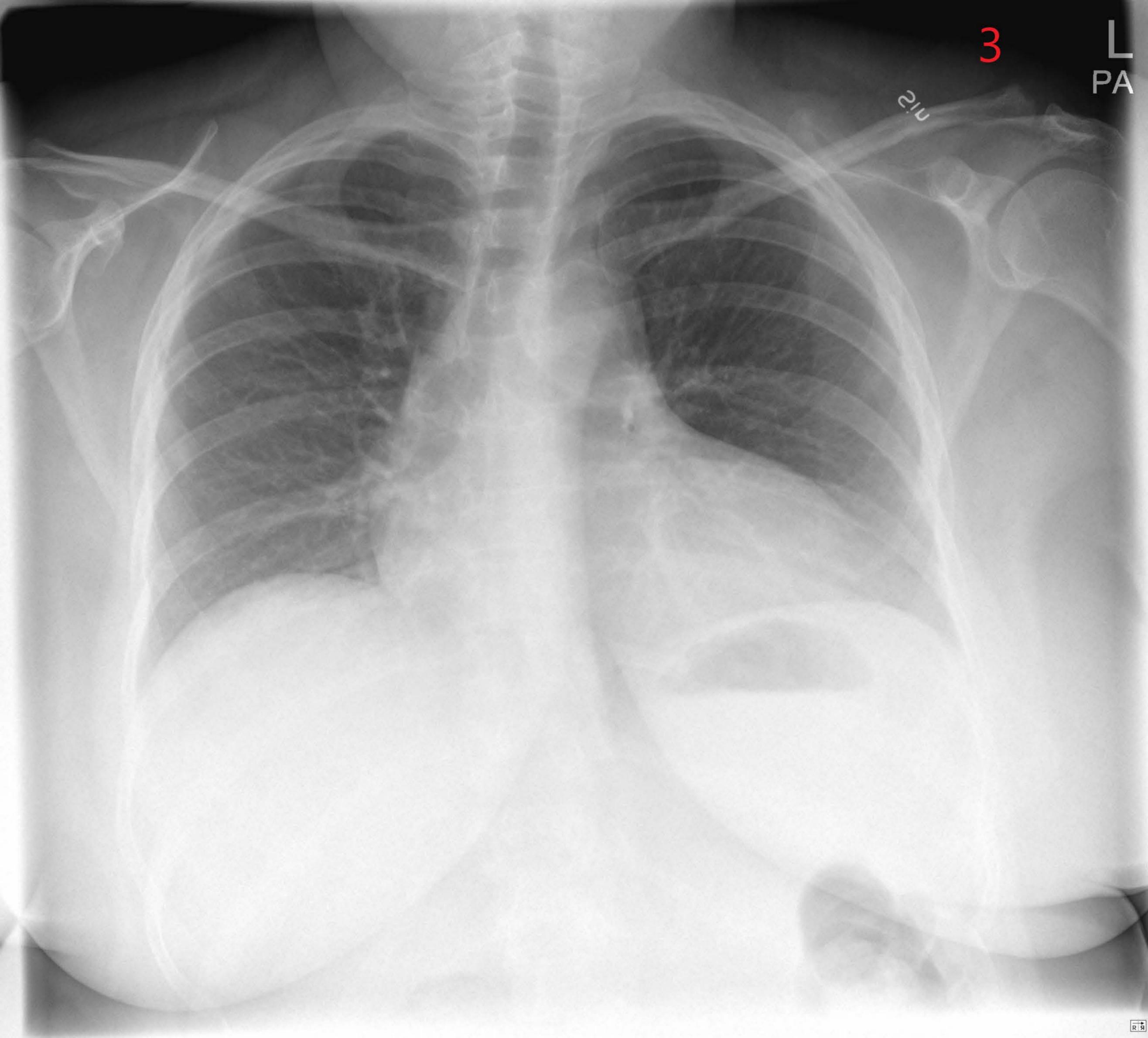

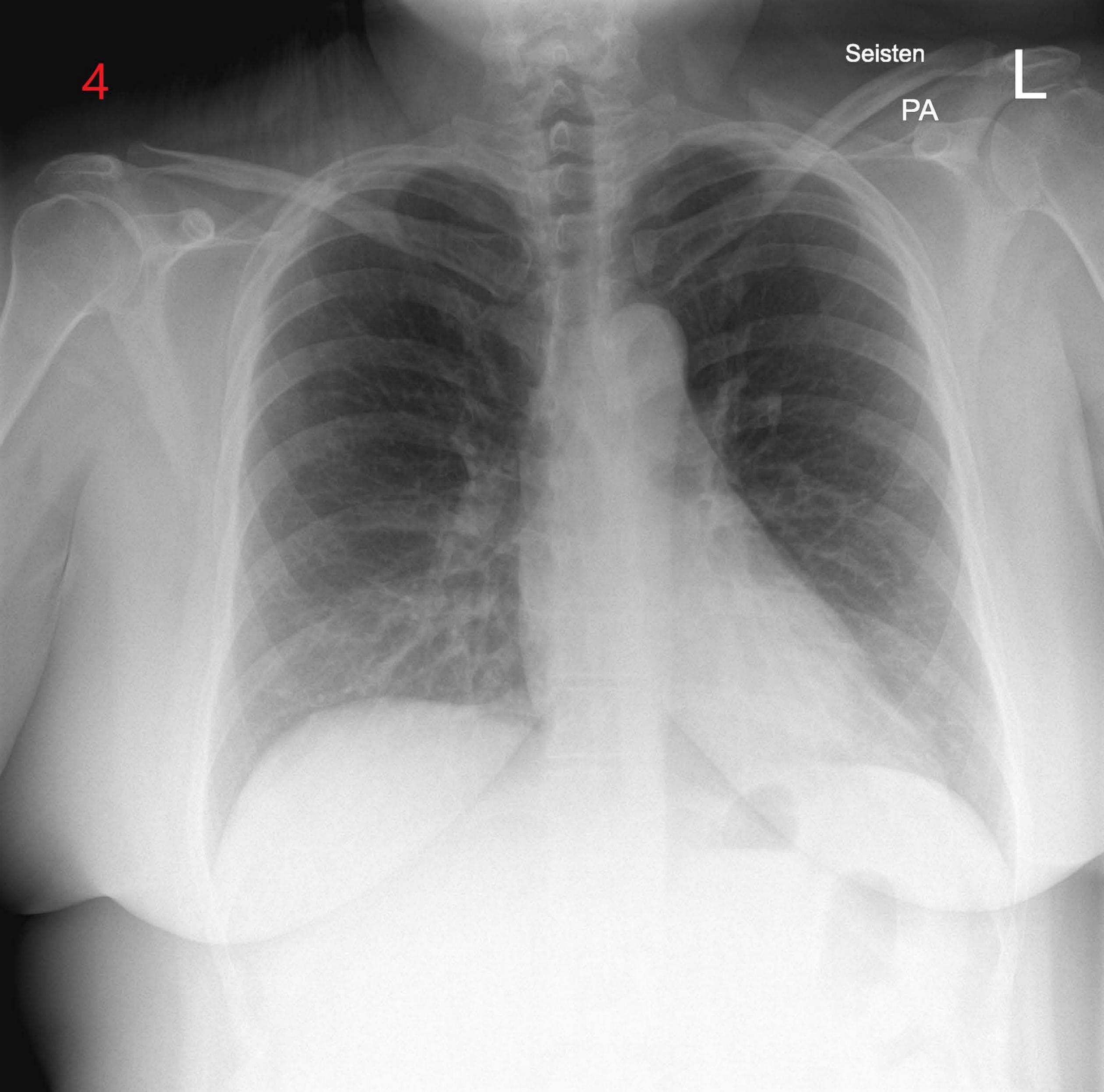

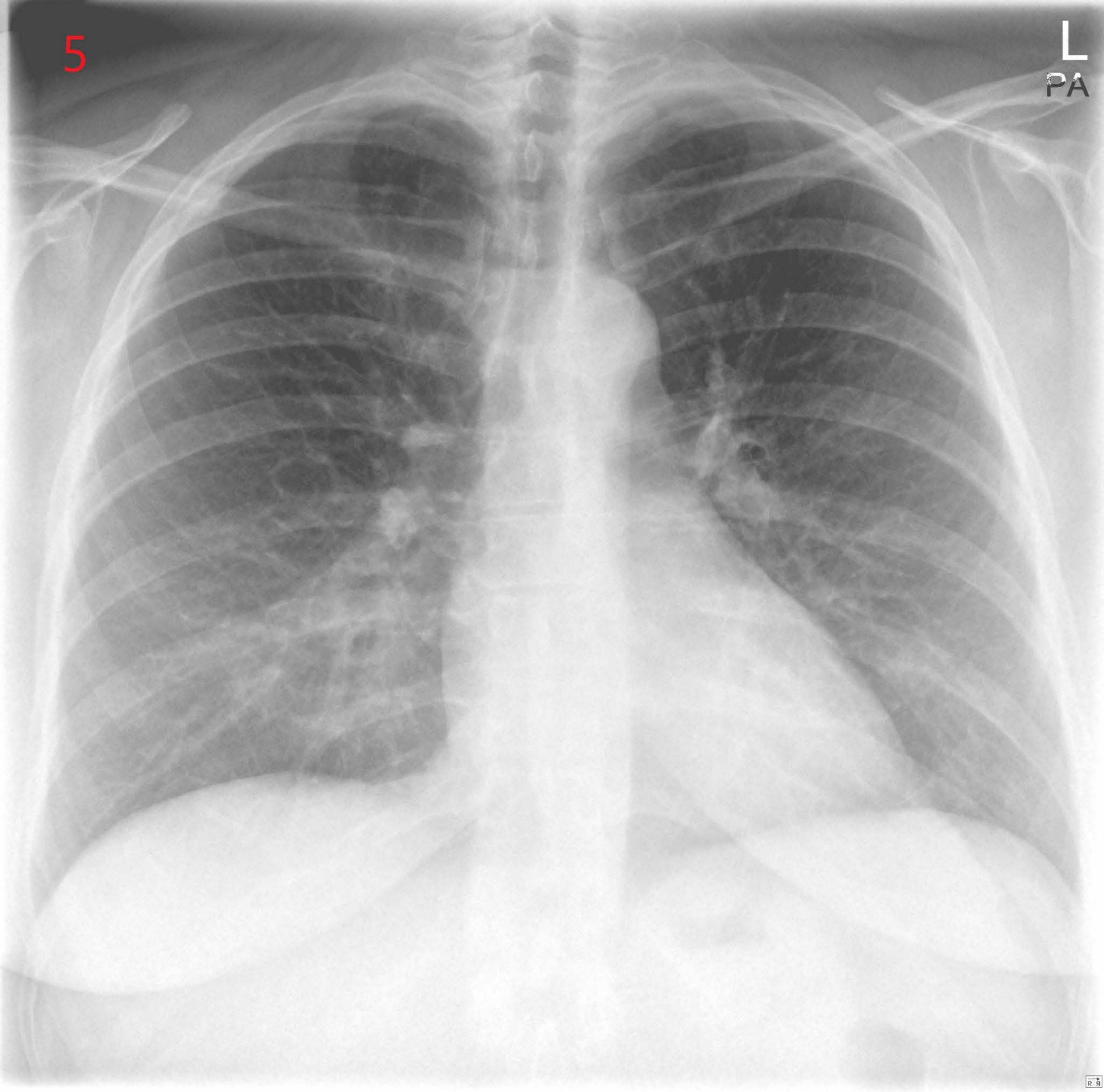

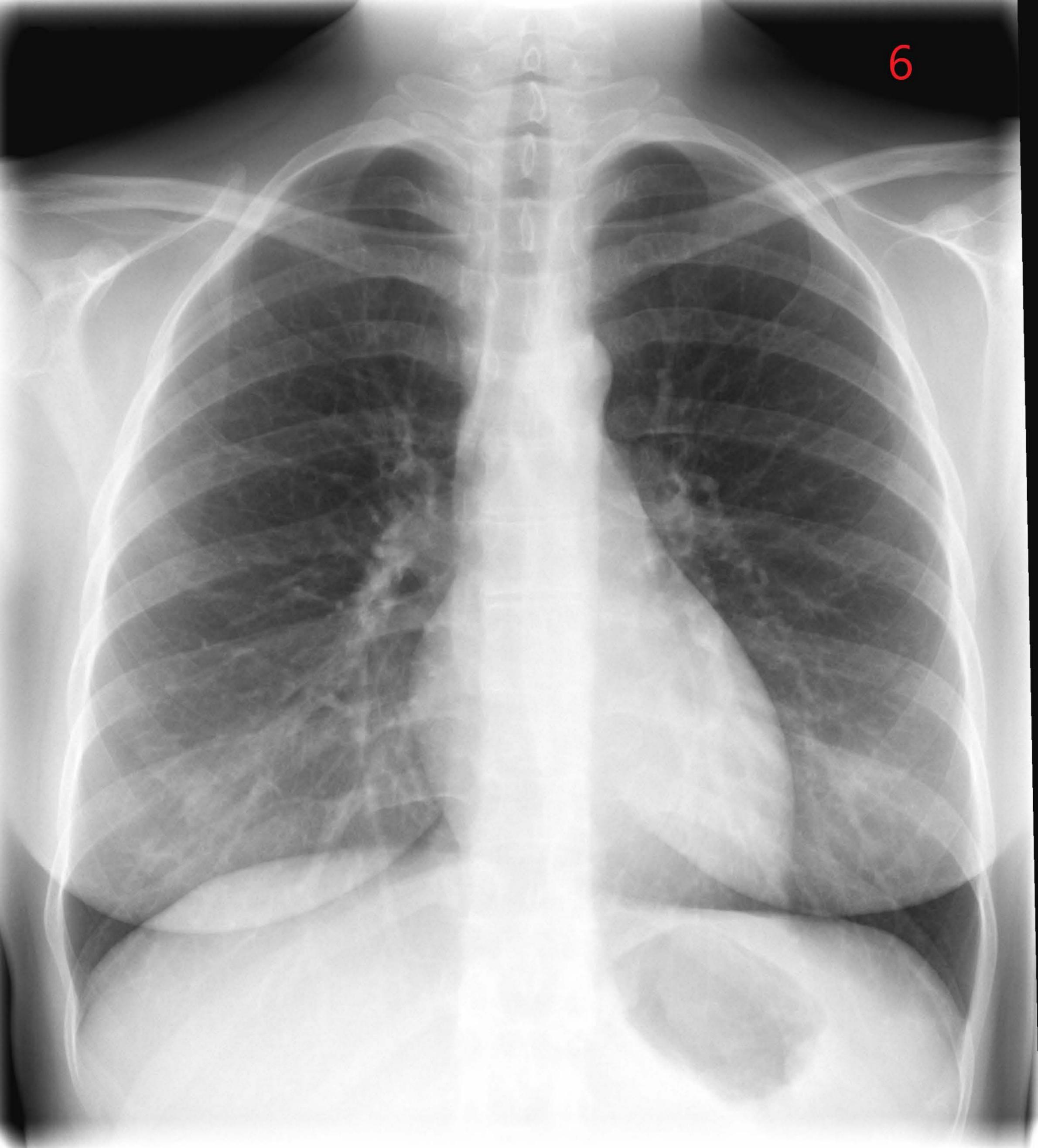

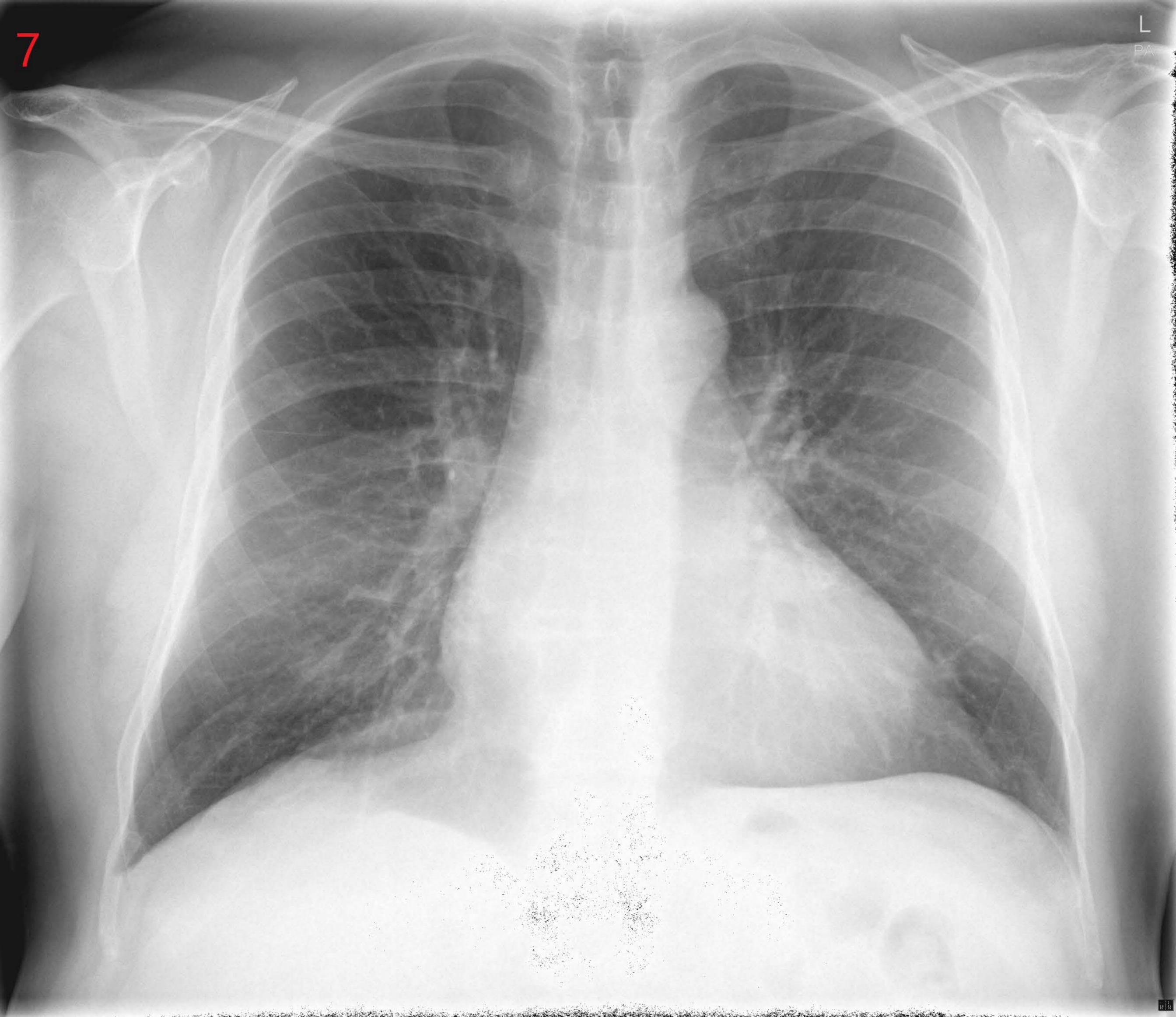

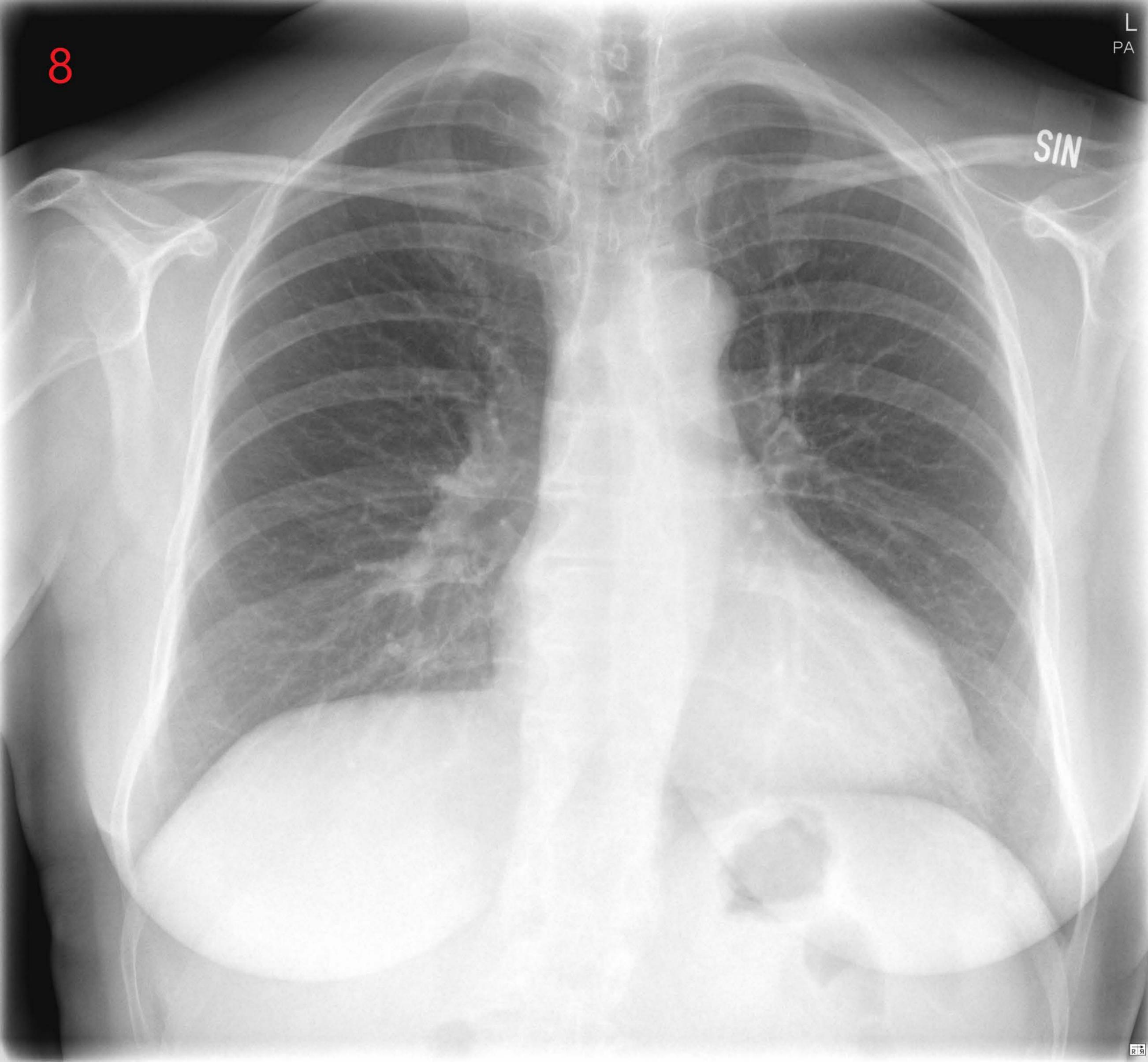

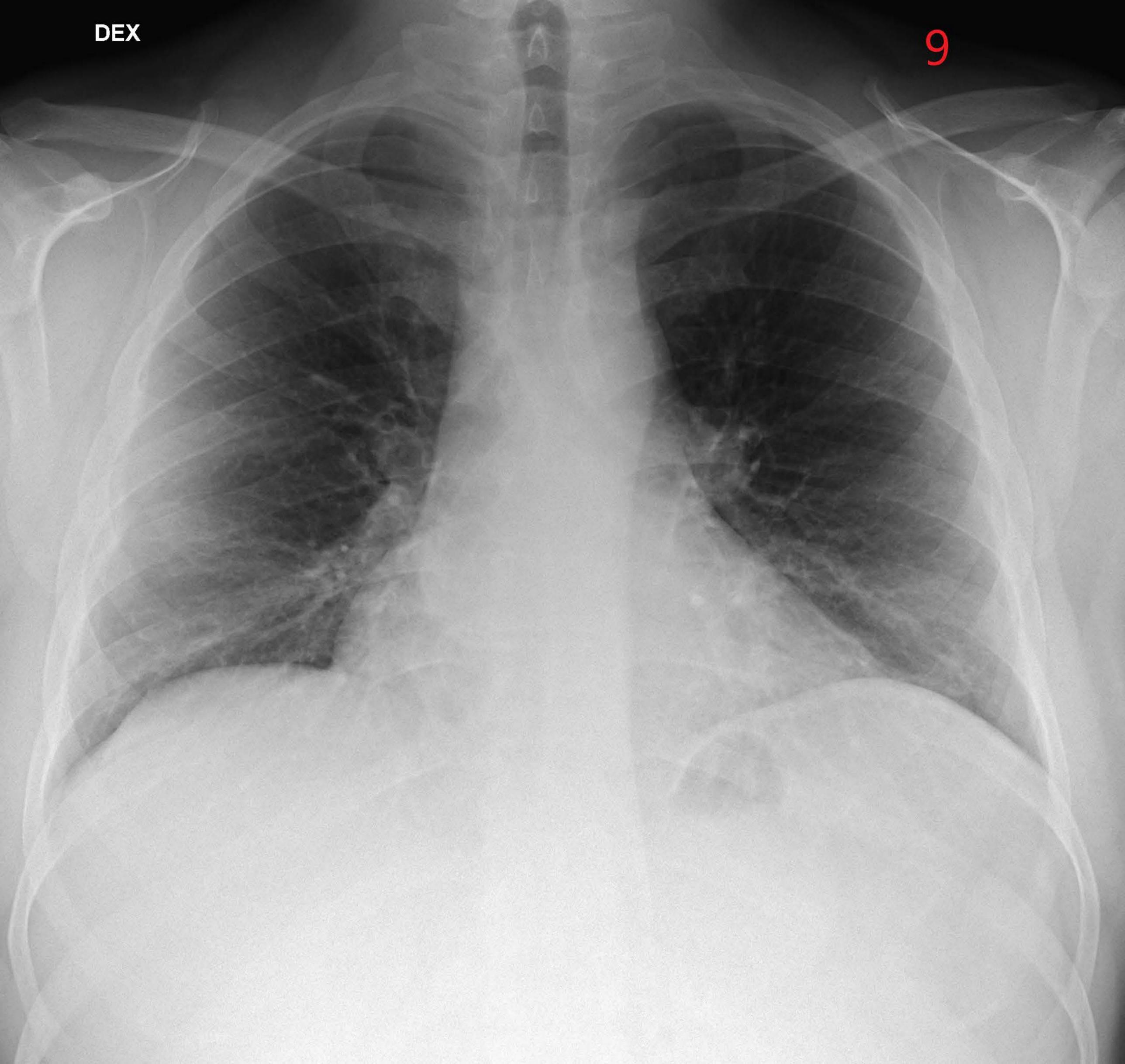

1- Left-sided pleural effusion

2- Slightly enlarged size of the heart

3- Slightly enlarged size of the heart

4- Increased left-sided pulmonary opacification

5- Increased right-sided pulmonary opacification

6- Increased bilateral basal pulmonary opacification

7- Slightly enlarged size of the heart

8- Slightly enlarged size of the heart

9- Increased bilateral basal pulmonary opacification